# {TikZ-Feynman}

## Feynman diagrams with TikZ

Version 1.0.0  19th January 2016

by Joshua Ellis

ARC Centre of Excellence for Particle Physics at the Terascale
School of Physics, The University of Melbourne VIC 3010, Australia

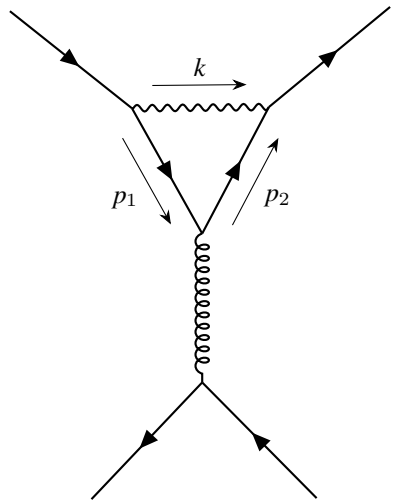

```
\feynmandiagram [large, vertical=e to f] {
  a -- [fermion] b -- [photon, momentum=\(k\)] c -- [fermion] d,
  b -- [fermion, momentum'=\(p_{1}\)] e -- [fermion, momentum'=\(p_{2}\)] c,
  e -- [gluon]  f,
  h -- [fermion] f -- [fermion] i;
};
```

## Contents

# Abstract

Ti*k*Z-Feynman is a LaTeX package allowing Feynman diagrams to be easily generated within LaTeX with minimal user instructions and without the need of external programs. It builds upon the Ti*k*Z package and leverages the graph placement algorithms from Ti*k*Z in order to automate the placement of many vertices. Ti*k*Z-Feynman still allows fine-tuned placement of vertices so that even complex diagrams can still be generated with ease.

# 1 Introduction

Ti*k*Z-Feynman provides a new way to draw Feynman diagrams in LaTeX that does not rely on external programs and uses a clear extensible syntax.

Feynman diagrams provide a description of interactions of subatomic particles in a form that is clearer and more succinct than the corresponding mathematical description. They were introduced by Feynman and first appear in his paper 'Space-Time Approach to Quantum Electrodynamics' [1]. Since then, LaTeX has become widely used to type-set scientific papers and currently, two leading methods of drawing Feynman diagrams in LaTeX are feynMF/feynMP [2] and AxoDraw [3], with the latter also featuring a GUI front-end called JaxoDraw [4, 5].

Both feynMF/feynMP and AxoDraw have quite complicated syntax. As an example, the code to generate an $\ell\ell \to \ell\ell$ scattering Feynman diagram in each package is:

AxoDraw
```
\begin{picture}
\ArrowLine(190,270)(160,300)
\ArrowLine(160,240)(190,270)
\ArrowLine(270,300)(240,270)
\ArrowLine(240,270)(270,240)
\Photon(190,270)(240,270){4}{4.5}
\Vertex(190,270){1.5}
\Vertex(240,270){1.5}
\end{picture}
```

feynMF/feynMP
```
\begin{fmfchar*}(40,30)
\fmfpen{thick}
\fmfleft{i1,i2}
\fmfright{o1,o2}
\fmf{fermion}{i1,v1,o1}
\fmf{fermion}{i2,v2,o2}
\fmf{photon,label=$q$}{v1,v2}
\fmfdot{v1,v2}
\end{fmfchar*}
```

Ti*k*Z-Feynman
```
\feynmandiagram [horizontal=a to b] {
  i1 -- [fermion] a -- [fermion] i2,
  a -- [photon] b,
  f1 -- [fermion] b -- [fermion] f2,
};
```

The learning curves for both AxoDraw and feynMF/feynMP are quite steep, and although this is partly resolved by using JaxoDraw, this requires an external program. In addition, the creation of Feynman diagrams in feynMF/feynMP requires further processing outside of LaTeX.

Ti*k*Z-Feynman on the other hand uses a clear syntax and delegates the positioning of vertices to algorithms originally developed by Hu [6], and Peter and Kozo [7], both of which were implemented into Ti*k*Z by Pohlmann [8]. Since Ti*k*Z-Feynman is built on Ti*k*Z, users can harness the power and extensibilty of Ti*k*Z [9] making it easy to extend to accommodate individual needs. In order to produce more complicated diagrams, relative or absolute positioning of vertices can also be used in Ti*k*Z-Feynman so that any diagram can be generated with relative ease.

Ti*k*Z-Feynman is made available through the Comprehensive TeX Archive Network (CTAN)[1] and can also be downloaded from the project page[2]. The project is open source and contributions are welcome. The management of bugs and feature requests is done at Github[3].

Ti*k*Z-Feynman's versioning will approximately follow semantic versioning. This means that changes in the third number (1.0.0 to 1.0.1) will consist of bug fixes and very minor changes but they should not change the output otherwise[4]. Changes in the second number (1.0.0 to 1.1.0) will consist of new features but everything should be backwards compatible. Finally, changes in the first number (1.0.0 to 2.0.0) indicates a major change in the package and code written for 1.0.0 is not guaranteed to work on 2.0.0. The intended version of this package to use should be indicated in the preamble with \tikzfeynmanset{compat=x.y.z} so the user may be informed of any discrepancy. If needed, earlier versions may be downloaded from the project page[1].

---

[1] https://ctan.org/pkg/tikz-feynman
[2] http://www.jpellis.me/projects/tikz-feynman
[3] https://github.com/JP-Ellis/tikz-feynman
[4] That is, with the exception of the bug that they are fixing.

## Licence

This *documentation* may be redistributed and/or modified under the terms of the gnu General Public License as published by the Free Software Foundation, either version 3 of the License, or (at your option) any later version.

The *code of this package* may be distributed and/or modified under the conditions of the LaTeX Project Public License, either version 1.3 of this license or (at your option) any later version.

This work has the LPPL maintenance status 'maintained'.

The Current Maintainer of this work is Joshua Ellis.

This package is distributed in the hope that it will be useful, but without any warranty; without even the implied warranty of merchantability or fitness for a particular purpose.

---

[1] http://www.jpellis.me/projects/tikz-feynman

# 2 Tutorial

## 2.1 Loading the Package

After installing the package, the Ti*k*Z-Feynman package can be loaded with `\usepackage{tikz-feynman}` in the preamble. It is recommend that you also place `\tikzfeynmanset{compat=1.0.0}` in the preamble to ensure that a new versions of Ti*k*Z-Feynman do not produce any undesirable changes without warning.

## 2.2 A First Diagram

Feynman diagrams can be declared with the `\feynmandiagram` command. It is analogous to the `\tikz` command from Ti*k*Z and requires a final semi-colon (;) to finish the environment. For example, a simple *s*-channel diagram is:

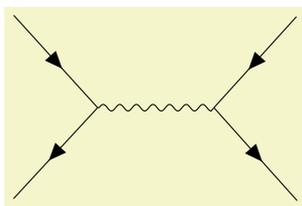

```
\feynmandiagram [horizontal=a to b] {
  i1 -- [fermion] a -- [fermion] i2,
  a -- [photon] b,
  f1 -- [fermion] b -- [fermion] f2,
};
```

Let's go through this example line by line:

LINE 1 `\feynmandiagram` introduces the Feynman diagram and allows for optional arguments to be given in the brackets [⟨options⟩]. In this instance, `horizontal=a to b` orients the algorithm outputs such that the line through vertices a and b is horizontal.

LINE 2 The left fermion line is drawn by declaring three vertices (i1, a and i2) and connecting them with edges `--`. Just like the `\feynmandiagram` command above, each edge also take optional arguments specified in brackets [⟨options⟩]. In this instance, we want these edges to have arrows to indicate that they are fermion lines, so we add the `fermion` style to them.

As you will see later on, optional arguments can also be given to the vertices in exactly the same way.

LINE 3 This edge connects vertices a and b with an edge styled as a photon. Since there is already a vertex labelled a, the algorithm will connect it to a new vertex labeled b.

LINE 4 This line is analogous to line 2 and introduces two new vertices, f1 and f2. It re-uses the previously labelled b vertex.

LINE 5 Finish the declaration of the Feynman diagram. The final semi-colon (;) is important.

The name given to each vertex in the graph does not matter. So in this example, i1, i2 denote the initial particles; f1, f2 denotes the final particles; and a, b are the end points of the propagator. The only important aspect is that what we called a in line 2 is also a in line 3 so that the underlying algorithm treats them as the same vertex.

The order in which vertices are declared does not matter as the default algorithm re-arranges everything[1]. For example, one might prefer to draw the fermion lines all at once, as with the following example (note also that the way we named vertices is completely different):

---
[1] It is possible for the algorithm to get a litte confused in some circumstances, but these cases should be rather rare. For some algorithms (such as the `layered layout`), the order in which vertices are introduces *does* matter. This is documented in section 3.2.2.

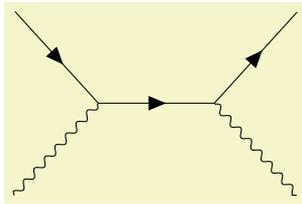

```
\feynmandiagram [horizontal=f2 to f3] {
  f1 -- [fermion] f2 -- [fermion] f3 -- [fermion] f4,
  f2 -- [photon] p1,
  f3 -- [photon] p2,
};
```

## 2.3 Adding Styles

So far, the examples have only used the `photon` and `fermion` styles. The Ti*k*Z-Feynman package comes with quite a few extra styles for edges and vertices which are all documented over in section 3. As an example, it is possible to add momentum arrows with `momentum=⟨text⟩`, and in the case of end vertices, the particle can be labelled with `particle=⟨text⟩`. As an example, we take the generic *s*-channel diagram from section 2.2 and make it a $e^+e^- \to \mu^+\mu^-$ diagram:

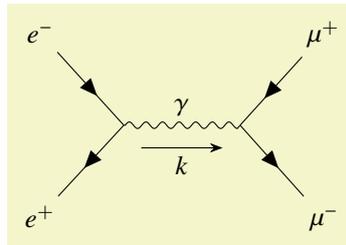

```
\feynmandiagram [horizontal=a to b] {
  i1 [particle=\(e^{-}\)] -- [fermion] a -- [fermion] i2 [particle=\(e^{+}\)],
  a -- [photon, edge label=\(\gamma\), momentum'=\(k\)] b,
  f1 [particle=\(\mu^{+}\)] -- [fermion] b -- [fermion] f2 [particle=\(\mu^{-}\)],
};
```

In addition to the style keys documented below, style keys from Ti*k*Z can be used as well:

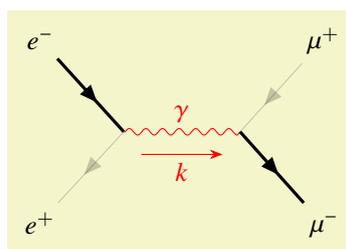

```
\feynmandiagram [horizontal=a to b] {
  i1 [particle=\(e^{-}\)] -- [fermion, very thick] a -- [fermion, opacity=0.2] i2 [particle=\(e^{+}\)],
  a -- [red, photon, edge label=\(\gamma\), momentum'={[arrow style=red]\(k\)}] b,
  f1 [particle=\(\mu^{+}\)] -- [fermion, opacity=0.2] b -- [fermion, very thick] f2 [particle=\(\mu^{-}\)],
};
```

For a list of all the various styles that Ti*k*Z provides, have a look at the Ti*k*Z manual; it is extremely thorough and provides many usage examples.

## 2.4 When the Algorithm Isn't Enough

By default, the \feynmandiagram and \diagram commands use the spring layout algorithm to place all the edges[1]. The spring layout algorithm attempts to 'spread out' the diagram as much as possible which—for most simpler diagrams—gives a satisfactory result; however in some cases, this does not produce the best diagram and this section will look at alternatives. There are three main alternatives:

Add invisible edges
: While still using the default algorithm, it is possible to force certain vertices to be closer together by adding extra edges and making them invisible through draw=none. The algorithm will treat these extra edges in the same way, but they are simply not drawn at the end;

Use a different algorithm
: In some circumstances, other algorithms may be better suited. Some of the other graph layout algorithms are listed in section 3.2.2, and an exhaustive list of all algorithms and their parameters is given in the Ti*k*Z manual;

Manual placement
: As a last resort, very complicated or unusual diagrams will require each vertex to be manually placed.

### 2.4.1 Invisible Edges

The underlying algorithm treats all edges in exactly the same way when calculating where to place all the vertices, and the actual drawing of the diagram (after the placements have been calculated) is done separately. Consequently, it is possible to add edges to the algorithm, but prevent them from being drawn by adding draw=none to the edge style.

This is particularly useful if you want to ensure that the initial or final states remain closer together than they would have otherwise as illustrated in the following example (note that opacity=0.2 is used instead of draw=none to illustrate where exactly the edge is located).

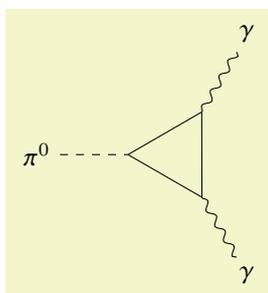

```
% No invisible to keep the two photons together
\feynmandiagram [small, horizontal=a to t1] {
  a [particle=\(\pi^{0}\)] -- [scalar] t1 -- t2 -- t3 -- t1,
  t2 -- [photon] p1 [particle=\(\gamma\)],
  t3 -- [photon] p2 [particle=\(\gamma\)],
};
```

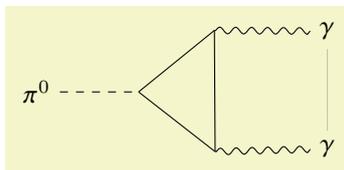

```
% Invisible edge ensures photons are parallel
\feynmandiagram [small, horizontal=a to t1] {
  a [particle=\(\pi^{0}\)] -- [scalar] t1 -- t2 -- t3 -- t1,
  t2 -- [photon] p1 [particle=\(\gamma\)],
  t3 -- [photon] p2 [particle=\(\gamma\)],
  p1 -- [opacity=0.2] p2,
};
```

---

[1] For more details on this layout and any other graph layouts available, see section 3.2.2

### 2.4.2 Alternative Algorithms

The graph drawing library from Ti*k*Z has several different algorithms to position the vertices[1] By default, \diagram and \feynmandiagram use the `spring layout` algorithm to place the vertices. The `spring layout` attempts to spread everything out as much as possible which, in most cases, gives a nice diagram; however, there are certain cases where this does not work. A good example where the `spring layout` doesn't work are decays where we have the decaying particle on the left and all the daughter particles on the right.

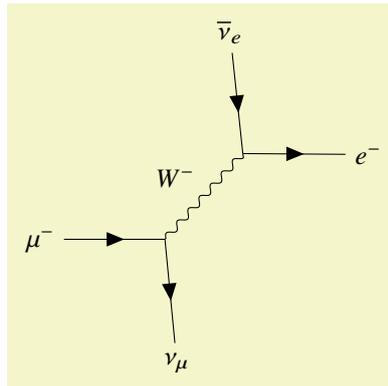

```
% Using the default spring layout
\feynmandiagram [horizontal=a to b] {
  a [particle=\(\mu^{-}\)] -- [fermion] b -- [fermion] f1 [particle=\(\nu_{\mu}\)],
  b -- [boson, edge label=\(W^{-}\)] c,
  f2 [particle=\(\overline \nu_{e}\)] -- [fermion] c -- [fermion] f3 [particle=\(e^{-}\)],
};
```

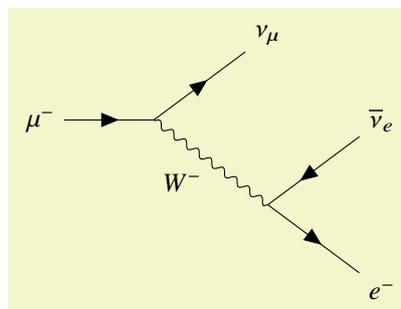

```
% Using the layered layout
\feynmandiagram [layered layout, horizontal=a to b] {
  a [particle=\(\mu^{-}\)] -- [fermion] b -- [fermion] f1 [particle=\(\nu_{\mu}\)],
  b -- [boson, edge label'=\(W^{-}\)] c,
  c -- [anti fermion] f2 [particle=\(\overline \nu_{e}\)],
  c -- [fermion] f3 [particle=\(e^{-}\)],
};
```

You may notice that in addition to adding the `layered layout` style to \feynmandiagram, we also changed the order in which we specify the vertices. This is because the `layered layout` algorithm does pay attention to the order in which vertices are declared (unlike the default `spring layout`); as a result, `c--f2, c--f3` has a different meaning to `f2--c--f3`. In the former case, f2 and f3 are both on the layer below c as desired; whilst the latter case places f2 on the layer above c (that, the same layer as where the $W^-$ originates).

---

[1] See section 3.2.2 for some alternative algorithms.

### 2.4.3 Manual Placement

In more complicated diagrams, it is quite likely that none of the algorithms work, no matter how many invisible edges are added. In such cases, the vertices have to be placed manually. Ti*k*Z-Feynman allows for vertices to be manually placed by using the \vertex command.

The \vertex command is available only within the feynman environment (which itself is only available inside a tikzpicture). The feynman environment loads all the relevant styles from Ti*k*Z-Feynman and declares additional Ti*k*Z-Feynman-specific commands such as \vertex and \diagram. This is inspired from PGFPlots and its use of the axis environment.

The \vertex command is very much analogous to the \node command from Ti*k*Z, with the notable exception that the vertex contents are optional; that is, you need not have {⟨*text*⟩} at the end. In the case where {} is specified, the vertex automatically is given the particle style, and otherwise it is a usual (zero-sized) vertex.

To specify where the vertices go, it is possible to give explicit coordinates though it is probably easiest to use the positioning library from Ti*k*Z which allows vertices to be placed relative to existing vertices[1]. By using relative placements, it is possible to easily tweak one part of the graph and everything will adjust accordingly—the alternative being to manually adjust the coordinates of every affected vertex.

Finally, once all the vertices have been specified, the \diagram* command is used to specify all the edges. This works in much the same way as \diagram (and also \feynmandiagram), except that it uses an very basic algorithm to place new nodes and allows existing (named) nodes to be included. In order to refer to an existing node, the node must be given in parentheses.

This whole process of specifying the nodes and then drawing the edges between them is shown below for the muon decay:

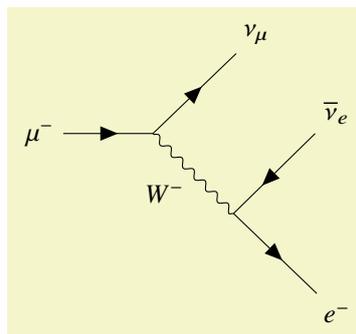

```
\begin{tikzpicture}
  \begin{feynman}
    \vertex (a) {\(\mu^{-}\)};
    \vertex [right=of a] (b);
    \vertex [above right=of b] (f1) {\(\nu_{\mu}\)};
    \vertex [below right=of b] (c);
    \vertex [above right=of c] (f2) {\(\overline \nu_{e}\)};
    \vertex [below right=of c] (f3) {\(e^{-}\)};

    \diagram* {
      (a) -- [fermion] (b) -- [fermion] (f1),
      (b) -- [boson, edge label'=\(W^{-}\)] (c),
      (c) -- [anti fermion] (f2),
      (c) -- [fermion] (f3),
    };
  \end{feynman}
\end{tikzpicture}
```

---

[1] The Ti*k*Z manual has some extensive documentation explaining how to use the positioning library.

# 3 Documentation

## 3.1 Commands & Environments

`\tikzfeynmanset{⟨options⟩}`

This command will process ⟨*options*⟩ using \pgfkeys with the default path set to /tikzfeynman. Typically, ⟨*options*⟩ will be a comma-separated list of the form ⟨*key*⟩=⟨*value*⟩, though the full power of the mechanism behind \pgfkeys can be used (see the TikZ manual for a complete description).

Typically, this is used in the preamble of the document to add or change certain keys for the whole document.

`\feynmandiagram[⟨TikZ options⟩][⟨diagram options⟩]{⟨diagram instructions⟩}`

This commands creates a {tikzpicture} and {feynman} environment, and places a \diagram inside with the provided ⟨*diagram instruction*⟩. Please refer to the documentation for \diagram for the ⟨*diagram instruction*⟩ syntax.

The optional arguments specified in ⟨*tikz options*⟩ are passed on to the {tikzpicture}, and the ⟨*diagram options*⟩ are passed on to \diagram. If only one optional argument is given, then the optional arguments are given to both. A single optional argument will usually suffice as most keys are recognized by both commands; however, in the event that a key is not recognized, both options are provided.

`\begin{feynman}[⟨options⟩]`
  ⟨*environment contents*⟩
`\end{feynman}`

The {feynman} environment is where all the drawing of Feynman diagrams takes place. It makes all the TikZ-Feynman styles available and defines commands such as \vertex and \diagram which are otherwise unavailable outside of this environment. The {feynman} environment is only accessible within the {tikzpicture} environment.

Options which are passed in ⟨*options*⟩ apply for the whole environment in the same way that the {scope} environment work in TikZ.

`\vertex[⟨options⟩] (⟨name⟩) at (⟨coordinate⟩) {⟨contents⟩};`

Defines a new vertex with the provided ⟨*name*⟩. If ⟨*contents*⟩ is not provided, the resulting vertex will have zero size. On the other hand, if ⟨*contents*⟩ is provided, the particle=⟨*contents*⟩ style is applied. Additional styles can be applied to the vertex through ⟨*options*⟩.

The final semicolon (;) is vital for this command since without it, the LaTeX engine will not know when the \vertex command ends. Additionally, this command *cannot* be chained like one can do with the inbuilt TikZ commands.

This command is only available with the {feynman} environment.

`\diagram[⟨options⟩]{⟨diagram instructions⟩}`

Begins a new diagram using the spring layout. Keys passed through ⟨*options*⟩ can include general TikZ keys, graph-specific keys and any applicable TikZ-Feynman keys too. Other algorithms (such as tree layout) can be passed through ⟨*options*⟩ and that will override the spring layout.

The syntax for the ⟨*diagram instructions*⟩ is thoroughly described in the TikZ manual, but in the context of this package, it will usually suffice to know the following:

- Vertices within the graph are specified with no delimiters (i.e. no parenthesis, no brackets) and only require spaces around either side. In order to refer to a vertex defined outside of the \diagram command, its name must be given in parenthesis: (⟨*name*⟩). Note that in order to refer to external vertices, one must use \diagram* as most algorithms (including the default spring layout) are incompatible with vertices defined outside of the algorithm.

When a vertex name is used multiple times, the underlying algorithm will consider them to be the same vertex and introduces additional edges.

Options can be given to the vertex in brackets after the name: ⟨*name*⟩ [⟨*options*⟩]. For vertices defined outside of the \diagram command, these options should be specified when the vertex is first declared.

- The edges between each pair of vertices is specified with --, and these can be chained together: a -- b -- c. In order to pass a style to the edge, it is specified in brackets after the dashed: --[⟨*options*⟩]. For example, to make on edge red, one would use --[red].
- A comma (,)—or equivalently a semicolon (;)—specifies the end of a sequence of edges and vertices and allows for another sequence to be started. So a -- b, c -- d will create two disconnected edges.
- Subgroups (aking to scopes in Ti*k*Z) are specified with braces: {[⟨*options*⟩]⟨*diagram instructions*⟩}. This can be quite useful when a lot of edges or nodes share a common style. For example, one could use {[edges={fermion}] a -- b -- c, x -- y -- z} and every edge will have the fermion style applied automatically.

Another useful feature of subgroups is that an edge to a group will create an edge to each vertex in that subgroup as shown below. The example also shows how they can be nested which in some cases (such as with a layered layout) can be extremely useful.

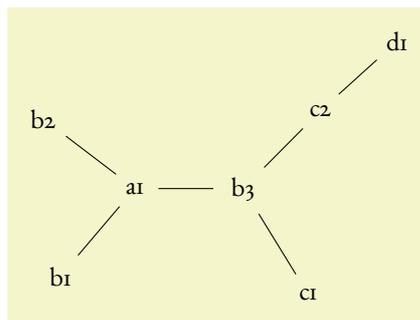

```
\feynmandiagram [nodes=circle, horizontal=a1 to b3] {
  a1 -- {b1, b2, b3 -- {c1, c2 -- d1}}
};
```

\diagram*[⟨*options*⟩]{⟨*diagram instructions*⟩}

Same as \diagram, but instead of using the spring layout algorithm to place the vertices, this uses the most basic algorithm. This basic algorithm in most cases will not produce a satisfactory diagram, but is intended to be used with vertices are declared and positioned outside of the \diagram* command. Essentially, \diagram* should be used only to connect existing vertices.

## 3.2 Keys & Styles

The various styles and options that allow the Feynman diagrams to be customized are defined in what Ti*k*Z calls *keys*. The documentation includes all keys which are defined within Ti*k*Z-Feynman which all begin with the prefix /tikzfeynman. In addition, a few of the keys from Ti*k*Z itself which are particularly useful to Ti*k*Z-Feynman are documented and these are prefixed with /tikz or /graph drawing. Please refer to the Ti*k*Z manual for a more in thorough documentation of the Ti*k*Z keys.

If you wish to modify the default TikZ-Feynman styles, the best way to do this is to use ⟨*key*⟩/.append style={...}. For example, to make every diagram red except for small diagrams which remain black, one would add to the preamble:

```
\tikzfeynmanset{
  every diagram/.append style={red},
  small/.append style={black},
}
```

If you are completely unhappy with a particular inbuilt style, you can define your own style with with ⟨*key*⟩/.style={...} as shown in the following example:

```
\tikzfeynmanset{
  myblob/.style={
    shape=circle,
    draw=blue,
    fill=red}
}
```

All the every ⟨*key*⟩ keys documented here are initially empty, so it is up to you whether you use ⟨*key*⟩/.append style or ⟨*key*⟩/.style. The predefined style keys (such as small, particle, fermion, etc.) should *never* by modified with ⟨*key*⟩/.style as that will overwrite the style entirely. Instead, modify the appropriate every ⟨*key*⟩ if available or use ⟨*key*⟩/.append style.

All the keys defined here are made available inside the {feynman} environment and inside \feynmandiagram; but if you wish to access them outside of this (say, in a regular {tikzpicture} environment), you will need to specify the full path with the leading /tikzfeynman.

### 3.2.1 Feynman Keys

/tikzfeynman/execute at begin feynman={⟨*T<sub>E</sub>X code*⟩}                                      (no default)
/tikzfeynman/execute at end feynman={⟨*T<sub>E</sub>X code*⟩}                                        (no default)

    Allows for custom code to be executed at the start or end of each {feynman} environment.

/tikzfeynman/every feynman                                                                            (style, no value)

    Set of styles which are applied to every {feynman} environments (and consequently, every apply inside all \feynmandiagram too). The style also applies to regular TikZ commands used inside the {feynman} environment.

```
\tikzfeynmanset{every feynman/.append style={red}}
\begin{tikzpicture}
  \node at (0, 0.5) {This is not red};
  \begin{feynman}
    \node at (0, -0.5) {This is red};
  \end{feynman}
\end{tikzpicture}
```

/tikzfeynman/inline=⟨*node*⟩                                                                          (style, no default)

    A style used to display a Feynman diagram inline (typically in an equation), and aligning such that its vertical placement is at the node specified. The node specification must enclosed in parentheses. For nodes which contain text (such as when the particle style is applied), it is possible to use the baseline of the text inside the

node to line up with the baseline of the equation by using (⟨*node*⟩.base) as demonstrated in the following example. Note that this key applies additional styles to make the diagram fit in an equation more nicely; if you do not wish to have these additional styles, use the `baseline` key.

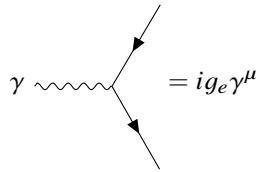

```
\begin{equation}
  \feynmandiagram [inline=(d.base), horizontal=d to b] {
    a -- [fermion] b -- [fermion] c,
    b -- [boson] d [particle=\(\gamma\)],
  };
  = i g_{e} \gamma^{\mu}
\end{equation}
```

/tikz/baseline=⟨*node*⟩ (no default)

Changes the vertical alignment of the Feynman diagram such that it diagram's baseline is at the node specified. This works in the same was as `inline=⟨node⟩`, but it does not apply any additional styles (notice how the following example is larger than the one above).

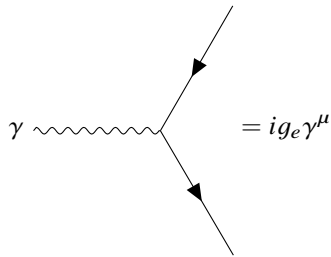

```
\begin{equation}
  \feynmandiagram [baseline=(d.base), horizontal=d to b] {
    a -- [fermion] b -- [fermion] c,
    b -- [boson] d [particle=\(\gamma\)],
  };
  = i g_{e} \gamma^{\mu}
\end{equation}
```

/graph drawing/horizontal=⟨*node*⟩ to ⟨*node*⟩ (no default)
/graph drawing/horizontal'=⟨*node*⟩ to ⟨*node*⟩ (no default)
/graph drawing/vertical=⟨*node*⟩ to ⟨*node*⟩ (no default)
/graph drawing/vertical'=⟨*node*⟩ to ⟨*node*⟩ (no default)

The underlying algorithm will arrange all the nodes relative to each other, but beyond that it has no idea how the overall graph should be oriented. By using one of the above keys, the final output of the algorithm is oriented and/or mirrored such that the two nodes specified are on the same horizontal (or vertical) line. The two nodes need not actually be connected by an edge for this to work.

The two ⟨*node*⟩ specifications should *not* be enclosed in parentheses, unlike the `inline` and `baseline` keys.

The `horizontal'` and `vertical'` keys work in the same was as `horizontal` and `vertical`, but with a flip.

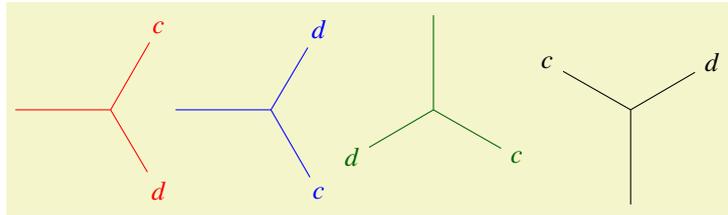

```
\feynmandiagram [inline=(b), horizontal=a to b, red] {
  a -- b -- {c [particle=\(c\)], d [particle=\(d\)]}
};
\feynmandiagram [inline=(b), horizontal'=a to b, blue] {
  a -- b -- {c [particle=\(c\)], d [particle=\(d\)]}
};
\feynmandiagram [inline=(b), vertical=a to b, green!40!black] {
  a -- b -- {c [particle=\(c\)], d [particle=\(d\)]}
};
\feynmandiagram [inline=(b), vertical=b to a, black] {
  a -- b -- {c [particle=\(c\)], d [particle=\(d\)]}
};
```

### 3.2.2 Diagram Keys

/tikzfeynman/every diagram (style, no value)

Set of styles which are applied to every diagram; that is, to everything inside the \feynmandiagram, \diagram and \diagram* commands but not the general {feynman} environment (see section 3.2.1 for that).

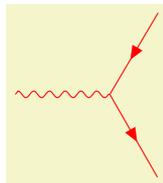

```
\tikzfeynmanset{every diagram/.append style={red}}
\feynmandiagram [small, horizontal=d to b] {
  a -- [fermion] b -- [fermion] c,
  b -- [boson] d,
};
```

/tikzfeynman/small (no value)
/tikzfeynman/medium (no value)
/tikzfeynman/large (no value)

Changes the default separation between the vertices and changes the size of arrows, blobs, and other shapes to fit different context. The small size is best used with when the diagram is quite simple and doesn't have too many annotations (such as momentum arrows and particle labels). The medium size is the default and is usually large enough that even diagrams with many labels and momentum arrows do not become too cluttered. Finally the large key is best for large illustrations as used on the title page of this document.

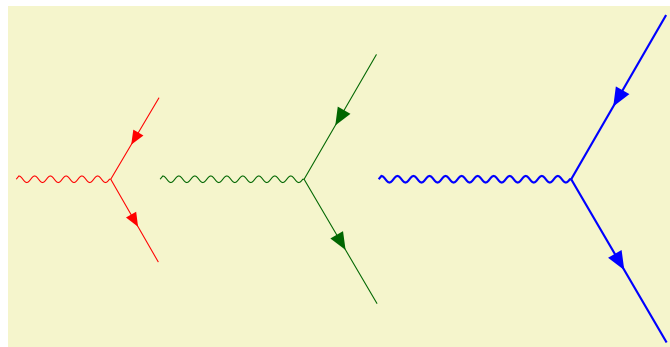

```
\feynmandiagram [baseline=(b), small, horizontal=d to b, red] {
  a -- [fermion] b -- [fermion] c,
  b -- [boson] d,
};
\feynmandiagram [baseline=(b), medium, horizontal=d to b, green!40!black] {
  a -- [fermion] b -- [fermion] c,
  b -- [boson] d,
};
\feynmandiagram [baseline=(b), large, horizontal=d to b, blue] {
  a -- [fermion] b -- [fermion] c,
  b -- [boson] d,
};
```

There are several algorithms which are available to place the vertices which are all provided within the `graph drawing` library from TikZ. Below are listed a few of these layouts which are more relevant for drawing Feynman diagrams. For a more complete description of how these algorithm work, please refer to the TikZ manual.

`/graph drawing/spring layout=`⟨*string*⟩ (no default)

Uses Hu's spring layout [6] as implemented by Pohlmann [8]. This models each edge as springs and attempts to spread everything out as much as possible. This is the default layout.

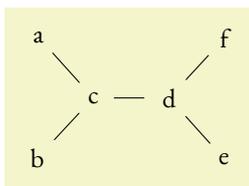

```
\feynmandiagram [nodes=circle, small, horizontal=c to d] {
  {a, b} -- c -- d -- {e, f},
};
```

`/graph drawing/spring electrical layout=`⟨*string*⟩ (no default)

Uses Hu's spring electrical layout [6] as implemented by Pohlmann [8]. This models each edge as springs and gives each vertex a charge. This algorithm allows for the charge of a particular vertex to be adjusted using the `electric charge` key (the default is 1).

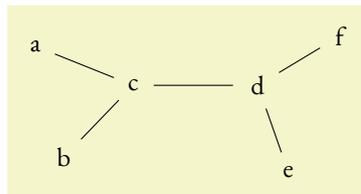

```
\feynmandiagram [nodes=circle,
    small, horizontal=c to d,
    spring electrical layout
] {
  {a, b [electric charge=2]} -- c -- d -- {e, f [electric charge=0.1]},
};
```

`/graph drawing/layered layout=`⟨*string*⟩ (no default)

Uses the Sugiyama layout algorithm [7] as implemented by Pohlmann [8] in order to place the node.

When an edge is specified, the first vertex is always located on the layer above the second vertex. This creates a hierarchy of vertices which is particularly useful for decays.

Two vertices can be forced to be on the same layer with the `/graph drawing/same layer` key.

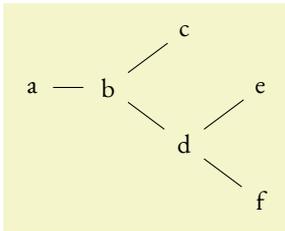
```
\feynmandiagram [nodes=circle, small, horizontal=a to b, layered layout] {
  a -- b -- {c, d -- {e, f}},
};
```

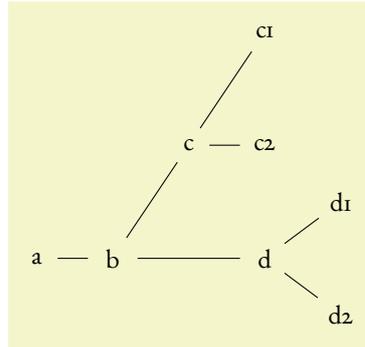

```
\feynmandiagram [nodes=circle, small, horizontal=a to b, layered layout] {
  a -- b -- {c -- {c1, c2}, d -- {d1, d2}},
  {[same layer] c1, d},
};
```

/graph drawing/tree layout=⟨*string*⟩ (no default)

Uses the Reingold–Tilform algorithm in order to place the node. This works in a similar way to the layered layout, but has quite a lot of additional options to handle missing children in the tree. Please refer to the Ti*k*Z manual for a thorough description of these additional features.

When an edge is specified, the first vertex is always located on the layer above the second vertex. This creates a hierarchy of vertices which is particularly useful for decays.

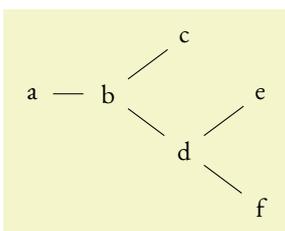
```
\feynmandiagram [nodes=circle, small, horizontal=a to b, tree layout] {
  a -- b -- {c, d -- {e, f}},
};
```

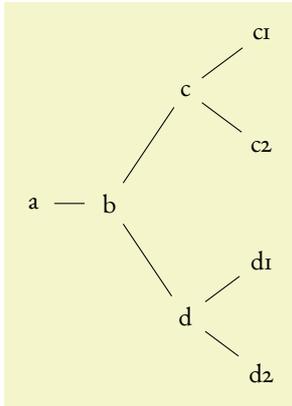
```
\feynmandiagram [nodes=circle, small, horizontal=a to b, tree layout] {
  a -- b -- {c -- {c1, c2}, d -- {d1, d2}},
};
```

/tikz/graphs/edges=⟨*options*⟩ (no default)
/tikz/graphs/nodes=⟨*options*⟩ (no default)

Just like it is possible to change the shape of every vertex or edge in the whole document, it is also possible to change the shape of every vertex or edge in a single diagram by modifying these keys and adding the desired styles.

### 3.2.3 Vertex Keys

/tikzfeynman/vertex (no value)

The default, base style applied to every vertex initially. Other styles are subsequently added. This sets the vertex shape to be a `coordinate`, that is, a null shape with no size or width.

/tikzfeynman/every ⟨*vertex shape*⟩ (style, initially empty)

The style of specific vertices can be modified by changing the appropriate every ⟨*vertex shape*⟩ key. For example, in order to change the style of every dot-styled vertex, you can modify the `every dot` key, or to modify every vertex globally, the `every vertex` key can be modified.

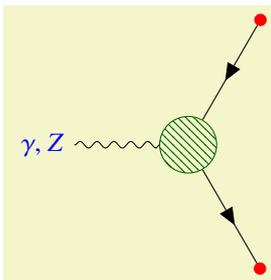
```
\tikzfeynmanset{
  every vertex/.style={red, dot},
  every particle/.style={blue},
  every blob/.style={draw=green!40!black, pattern color=green!40!black},
}
\feynmandiagram [horizontal=a to b] {
  a [particle={\(\gamma, Z\)}] -- [boson] b [blob],
  c -- [fermion] b -- [fermion] d,
};
```

/tikzfeynman/dot (no value)

Modifies the vertex so that it has a small filled circle.

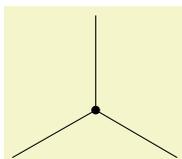
```
\feynmandiagram [small] {
  a -- b [dot] -- {c, d}
};
```

`/tikzfeynman/square dot` (no value)

Modifies the vertex so that it has a small filled square.

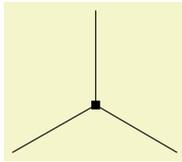
```
\feynmandiagram [small] {
  a -- b [square dot] -- {c, d}
};
```

`/tikzfeynman/empty dot` (no value)

Modifies the vertex so that it has a small empty circle.

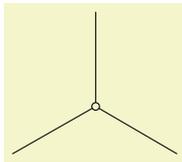
```
\feynmandiagram [small] {
  a -- b [empty dot] -- {c, d}
};
```

`/tikzfeynman/crossed dot` (no value)

Modifies the vertex so that it has a small circle with a cross inside.

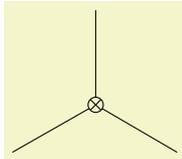
```
\feynmandiagram [small] {
  a -- b [crossed dot] -- {c, d}
};
```

`/tikzfeynman/blob` (no value)

Modifies the vertex so that it is a large blob, usually used to denote an effective operator.

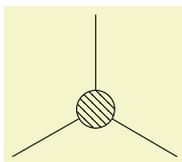
```
\feynmandiagram [small] {
  a -- b [blob] -- {c, d}
};
```

`/tikzfeynman/particle`=⟨*name*⟩ (no default)

Modifies the vertex so that it shows ⟨*name*⟩. This is intended to label initial and final particles, but it should not be used on internal vertices as it will result in the lines at the vertex having a gap. For propagators (the `edge label` key is much more appropriate).

Note that if ⟨*name*⟩ contains characters such as brackets (`[]`) or commas (`,`), the whole ⟨*name*⟩ has to be enclosed in braces (`{}`); otherwise, the parser will interpret the comma as the end of the ⟨*name*⟩ and the start of the next key, or the closing bracket as the end of all optional arguments.

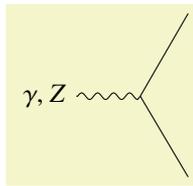
```
\feynmandiagram [small, horizontal=a to b] {
  a [particle={\(\gamma, Z\)}] -- [boson] b -- {c, d},
};
```

### 3.2.4 Edge Keys

Just like with the various vertex keys, each edge type has a corresponding every ⟨*edge type*⟩; however, due to the existence of very similar keys such as `scalar`, `charged scalar` and `anti charged scalar`, more specific keys inherit styles from less specific ones. For example, styles in every `charged scalar` will apply to `charged scalar` and `anti charged scalar` but not `scalar` whilst styles in every `scalar` will apply to all three.

`/tikzfeynman/every edge` (style, initially empty)

  A style to apply to every edge initially.

`/tikzfeynman/every ⟨edge style⟩` (style, initially empty)

  The style of specific edges can be modified by changing the appropriate every ⟨*edge style*⟩ key. For example, in order to make a global change to every boson, you can modify the every boson key.

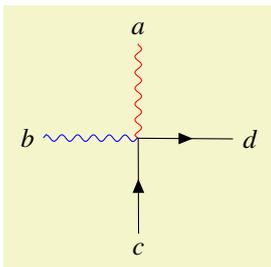
```
\tikzfeynmanset{
  every edge/.style={fermion},
  every boson/.style={red},
  every photon/.style={blue},
}
\feynmandiagram [small] {
  a [particle=\(a\)] -- [boson] o -- [photon] b [particle=\(b\)],
  f1 [particle=\(c\)] -- o -- f2 [particle=\(d\)],
};
```

`/tikzfeynman/boson` (no value)

Draws a sinusoidal line to denote a boson.

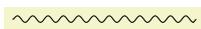
```
\feynmandiagram [horizontal=a to b] {a -- [boson] b};
```

`/tikzfeynman/charged boson` (no value)

Draws a sinusoidal line with an arrow to denote a charged boson.

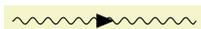
```
\feynmandiagram [horizontal=a to b] {a -- [charged boson] b};
```

`/tikzfeynman/anti charged boson` (no value)

Draws a sinusoidal line with an arrow pointing the other way to to denote a anti charged boson.

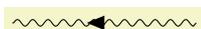
```
\feynmandiagram [horizontal=a to b] {a -- [anti charged boson] b};
```

/tikzfeynman/photon (no value)
Draws a sinusoidal line to denote a photon.

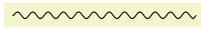 `\feynmandiagram [horizontal=a to b] {a -- [photon] b};`

/tikzfeynman/scalar (no value)
Draws a dashed line to denote a scalar.

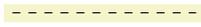 `\feynmandiagram [horizontal=a to b] {a -- [scalar] b};`

/tikzfeynman/charged scalar (no value)
Draws a dashed line with an arrow to denote a charged scalar.

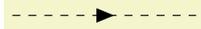 `\feynmandiagram [horizontal=a to b] {a -- [charged scalar] b};`

/tikzfeynman/anti charged scalar (no value)
Draws a dashed line with an arrow pointing the other way to denote a charged scalar antiparticle.

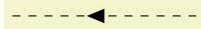 `\feynmandiagram [horizontal=a to b] {a -- [anti charged scalar] b};`

/tikzfeynman/ghost (no value)
Draws a dotted line to denote a ghost.

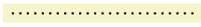 `\feynmandiagram [horizontal=a to b] {a -- [ghost] b};`

/tikzfeynman/fermion (no value)
Draws a solid line with an arrow to denote a fermion.

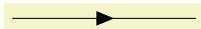 `\feynmandiagram [horizontal=a to b] {a -- [fermion] b};`

/tikzfeynman/anti fermion (no value)
Draws a solid line with an arrow pointing the other way to denote an antifermion.

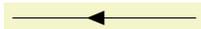 `\feynmandiagram [horizontal=a to b] {a -- [anti fermion] b};`

/tikzfeynman/majorana (no value)
Draws a solid line with two arrows pointing to the center to denote an Majorana particle.

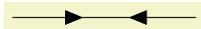 `\feynmandiagram [horizontal=a to b] {a -- [majorana] b};`

`/tikzfeynman/anti majorana` (no value)

Draws a solid line with two arrows pointing to the ends to denote a Majorana particle.

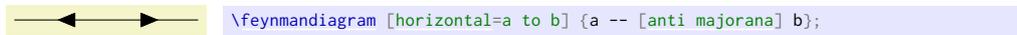 `\feynmandiagram [horizontal=a to b] {a -- [anti majorana] b};`

`/tikzfeynman/gluon` (no value)

Draws a coiled line to denote a gluon.

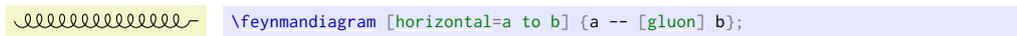 `\feynmandiagram [horizontal=a to b] {a -- [gluon] b};`

`/tikz/edge label=`⟨*text*⟩ (no default)
`/tikz/edge label'=`⟨*text*⟩ (no default)

Places a label halfway along the edge with the given text. The primed key switches which side of the edge the label is placed.

`/tikzfeynman/insertion=[`⟨*options*⟩`]`⟨*distance*⟩ (no default)

Places an insertion (for mass or momentum insertion) along an edge. The distance specifies how far along the edge the insertion should be placed such that 0 and 1 respectively correspond to the start and the end of the edge.

Multiple insertions can be placed along a single edge by repeating the style key.

Through the ⟨*options*⟩ argument, the insertion size and style can be changed.

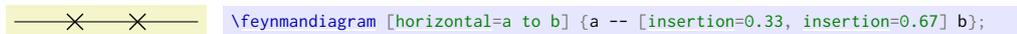 `\feynmandiagram [horizontal=a to b] {a -- [insertion=0.33, insertion=0.67] b};`

`/tikzfeynman/insertion/size=`⟨*distance*⟩ (no default, initially 3pt)

Specifies how big the insertion should be. The length of each edge starting from the center will be $\sqrt{2} \times$ ⟨*distance*⟩.

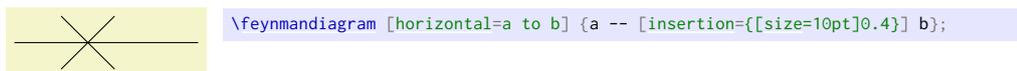 `\feynmandiagram [horizontal=a to b] {a -- [insertion={[size=10pt]0.4}] b};`

`/tikzfeynman/insertion/style=`⟨*distance*⟩ (no default, initially empty)

Specifies additional styles to applying to the lines of the insertion.

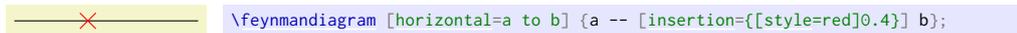 `\feynmandiagram [horizontal=a to b] {a -- [insertion={[style=red]0.4}] b};`

### 3.2.5 Momentum Keys

`/tikzfeynman/momentum=[`⟨*options*⟩`]`⟨*text*⟩ (no default)
`/tikzfeynman/momentum'=[`⟨*options*⟩`]`⟨*text*⟩ (no default)
`/tikzfeynman/reversed momentum=[`⟨*options*⟩`]`⟨*text*⟩ (no default)
`/tikzfeynman/reversed momentum'=[`⟨*options*⟩`]`⟨*text*⟩ (no default)
`/tikzfeynman/rmomentum=[`⟨*options*⟩`]`⟨*text*⟩ (no default)

`/tikzfeynman/rmomentum'=[⟨options⟩]⟨text⟩` (no default)

Places a momentum arrow on the specified edge with label given by ⟨*text*⟩. The primed (`'`) version place the momentum arrow on the other side of the edge; that is, if the momentum arrow was on the right, it will be placed on the left of the edge. The `reversed momentum` and `reversed momentum'` keys are analogous to `momentum` and `momentum'` except that the momentum arrow points in the opposite direction. Finally, the `rmomentum` and `rmomentum'` are aliases of `reversed momentum` and `reversed momentum'`.

Note that due to the way the arrow is drawn, it doesn't inherit styles of the edge. As a result, they have to be re-specified through ⟨*options*⟩.

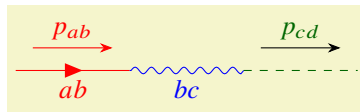

```
\feynmandiagram [layered layout, horizontal=a to b] {
  a -- [red, fermion, edge label'=\(ab\), momentum={[arrow style=red]\(p_{ab}\)}] b
    -- [blue, photon, edge label'=\(bc\)] c
    -- [green!40!black, scalar, momentum=\(p_{cd}\)] d,
};
```

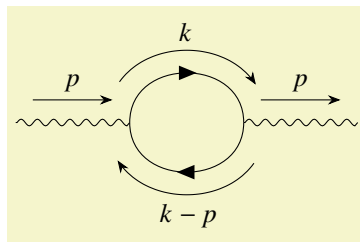

```
\feynmandiagram [layered layout, horizontal=b to c] {
  a -- [photon, momentum=\(p\)] b
    -- [fermion, half left, looseness=1.5, momentum=\(k\)] c
    -- [fermion, half left, looseness=1.5, momentum=\(k-p\)] b,
  c -- [photon, momentum=\(p\)] d,
};
```

The ⟨*options*⟩ allows for the following options to customize the appearance of the momentum arrows. These can be either set globally with the usual `\tikzfeynmanset`, or can be set on a individual basis through the ⟨*options*⟩ argument of the `momentum` key. In the latter case, only the ⟨*key*⟩ in `/tikzfeynman/momentum/`⟨*key*⟩ is required.

`/tikzfeynman/momentum/label distance=`⟨*distance*⟩ (no default, initially `0pt`)

Set the separation between the text and the arrow. Note that the text is still surrounded by an `inner sep=0.3333em` by default so the default distance of `0pt` will not result in the momentum label touching the arrow.

`/tikzfeynman/momentum/arrow distance=`⟨*distance*⟩ (no default, initially `3mm`)

Set the separation between the edge and the arrow.

`/tikzfeynman/momentum/arrow shorten=`⟨*distance*⟩ (no default, initially `0.15`)

Specify the fraction of the total edge length by which the momentum arrow is shortened at each end.

`/tikzfeynman/momentum/label style=`⟨*style*⟩ (no default, initially empty)

Define styles to apply to the momentum label node.

`/tikzfeynman/momentum/arrow style`=⟨*style*⟩ (no default, initially empty)

Define style to apply to the momentum arrow.

### 3.2.6 Modifier Keys

Modifier keys serve only to slightly modify a small feature of the edge.

`/tikzfeynman/half left` (no value)
`/tikzfeynman/half right` (no value)
`/tikzfeynman/quarter left` (no value)
`/tikzfeynman/quarter right` (no value)

Modifies the edge so that it bends left or right in such a way that it completes half a circle, or a quarter of a circle.

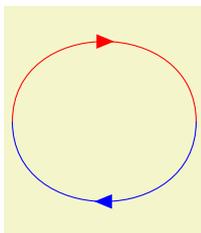
```
\feynmandiagram [horizontal=a to b] {
  a -- [red, fermion, half left] b -- [blue, fermion, half left] a,
};
```

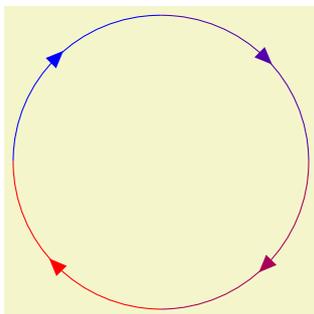
```
\feynmandiagram [horizontal=a to c] {
  a -- [red!0!blue, fermion, quarter left] b
    -- [red!33!blue, fermion, quarter left] c
    -- [red!66!blue, fermion, quarter left] d
    -- [red!100!blue, fermion, quarter left] a,
};
```

`/tikz/out`=⟨*angle*⟩ (no default)
`/tikz/in`=⟨*angle*⟩ (no default)

Specifies the angle at which the edge leaves the first vertex in an edge and the angle at which it enters the second vertex in an edge.

`/tikz/relative`=⟨*true or false*⟩ (default `true`)

If `relative` is set to `false`, the angle is relative to the paper whilst when `relative` is set to `true`, the angle is relative to the straight line joining the two vertices.

`/tikz/looseness`=⟨*number*⟩ (no default, initially 1)

As the name suggests, this specifies how 'loose' or 'tight' a curve is connecting two vertices.

# 4 Examples

Below are a few diagrams which demonstrate how the package can be used in some more practical examples..

## Vertex Rule

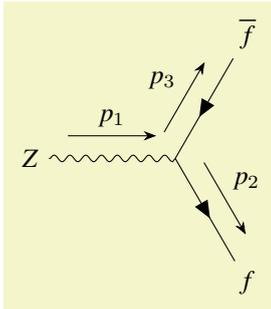

```
\feynmandiagram [horizontal=a to b] {
  a [particle=\(Z\)] -- [photon, momentum=\(p_{1}\)] b,
  f1 [particle=\(\overline f\)]
      -- [fermion, rmomentum'=\(p_{3}\)] b
      -- [fermion, momentum=\(p_{2}\)] f2 [particle=\(f\)],
};
```

## Tree Level Diagrams

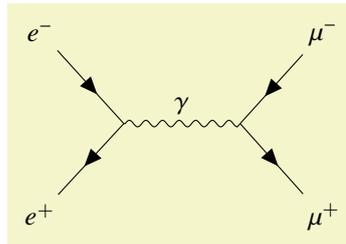

```
\feynmandiagram [horizontal=a to b] {
  i1 [particle=\(e^{-}\)] -- [fermion] a -- [fermion] i2 [particle=\(e^{+}\)],
  a -- [photon, edge label=\(\gamma\)] b,
  f1 [particle=\(\mu^{-}\)] -- [fermion] b -- [fermion] f2 [particle=\(\mu^{+}\)],
};
```

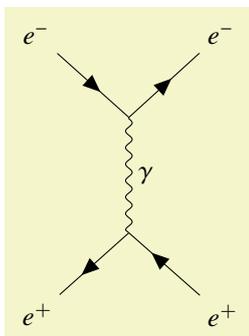

```
\feynmandiagram [vertical'=a to b] {
  i1 [particle=\(e^{-}\)]
      -- [fermion] a
      -- [fermion] f1 [particle=\(e^{-}\)],
  a -- [photon, edge label=\(\gamma\)] b,
  i2 [particle=\(e^{+}\)]
      -- [anti fermion] b
      -- [anti fermion] f2 [particle=\(e^{+}\)],
};
```

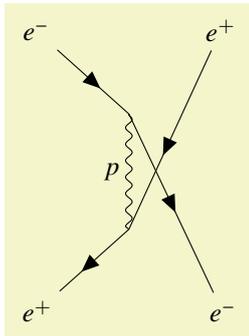
```
\begin{tikzpicture}
  \begin{feynman}
    \diagram [vertical'=a to b] {
      i1 [particle=\(e^{-}\)]
        -- [fermion] a
        -- [draw=none] f1 [particle=\(e^{+}\)],
      a -- [photon, edge label'=\(p\)] b,
      i2 [particle=\(e^{+}\)]
        -- [anti fermion] b
        -- [draw=none] f2 [particle=\(e^{-}\)],
    };
    \diagram* {
      (a) -- [fermion] (f2),
      (b) -- [anti fermion] (f1),
    };
  \end{feynman}
\end{tikzpicture}
```

## Loops

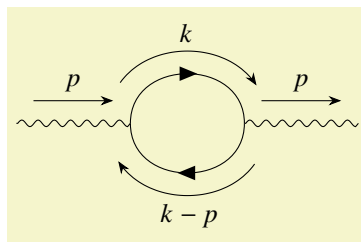

```
\feynmandiagram [layered layout, horizontal=b to c] {
  a -- [photon, momentum=\(p\)] b
    -- [fermion, half left, momentum=\(k\)] c
    -- [fermion, half left, momentum=\(k-p\)] b,
  c -- [photon, momentum=\(p\)] d,
};
```

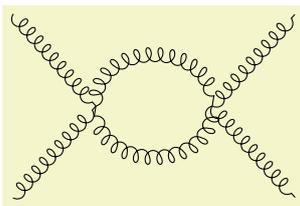

```
\feynmandiagram [layered layout, horizontal=a to b] [edges=gluon] {
  {i1, i2} -- a -- [half left] b -- [half left] a,
  b -- {f1, f2},
};
```

Box Diagrams

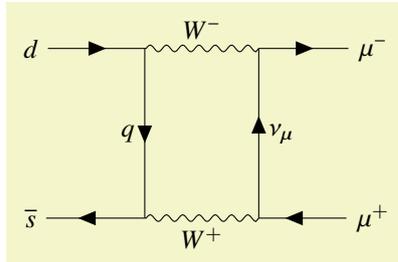

```
\feynmandiagram [layered layout, horizontal=a to b] {
  % Draw the top and bottom lines
  i1 [particle=\(d\)]
    -- [fermion] a
    -- [photon, edge label=\(W^{-}\)] b
    -- [fermion] f1 [particle=\(\mu^{-}\)],
  i2 [particle=\(\overline s\)]
    -- [anti fermion] c
    -- [photon, edge label'=\(W^{+}\)] d
    -- [anti fermion] f2 [particle=\(\mu^{+}\)],
  % Draw the two internal fermion lines
  { [same layer] a -- [fermion,      edge label'=\(q\)] c },
  { [same layer] b -- [anti fermion, edge label=\(\nu_{\mu}\)] d},
};
```

## Meson decay and mixing

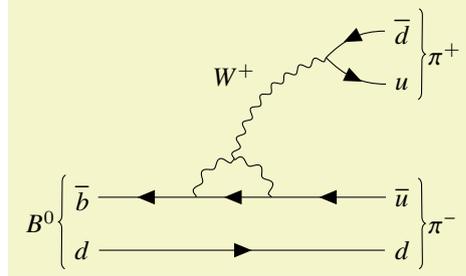

```
\begin{tikzpicture}
  \begin{feynman}
    \vertex (a1) {\(\overline b\)};
    \vertex[right=1.5cm of a1] (a2);
    \vertex[right=1cm of a2] (a3);
    \vertex[right=1.5cm of a3] (a4) {\(\overline u\)};

    \vertex[below=2em of a1] (b1) {\(d\)};
    \vertex[below=2em of a4] (b2) {\(d\)};

    %% See section 13.5 of PGF/TikZ manual
    \vertex at ($(a2)!0.5!(a3)!0.5cm!90:(a3)$) (d);
    %% Equivalent way to obtain (d):
    % \vertex at ($(b2)!0.5!(b3) + (0, -0.5cm)$) (d);
    \vertex[above=of a4] (c1) {\(u\)};
    \vertex[above=2em of c1] (c3) {\(\overline d\)};
    \vertex at ($(c1)!0.5!(c3) - (1cm, 0)$) (c2);

    \diagram* {
      (a4) -- [fermion] (a3) -- [fermion] (a2) -- [fermion] (a1),
      (b1) -- [fermion] (b2),
      (c3) -- [fermion, out=180, in=45] (c2) -- [fermion, out=-45, in=180] (c1),
      (a2) -- [boson, quarter left] (d) -- [boson, quarter left] (a3),
      (d) -- [boson, bend left, edge label=\(W^{+}\)] (c2),
    };

    \draw [decoration={brace}, decorate] (b1.south west) -- (a1.north west)
          node [pos=0.5, left] {\(B^{0}\)};
    \draw [decoration={brace}, decorate] (c3.north east) -- (c1.south east)
          node [pos=0.5, right] {\(\pi^{+}\)};
    \draw [decoration={brace}, decorate] (a4.north east) -- (b2.south east)
          node [pos=0.5, right] {\(\pi^{-}\)};
  \end{feynman}
\end{tikzpicture}
```

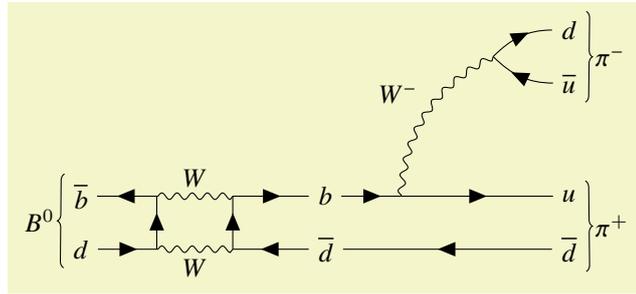

```
\begin{tikzpicture}
  \begin{feynman}
    \vertex (a1) {\(\overline b\)};
    \vertex[right=1cm of a1] (a2);
    \vertex[right=1cm of a2] (a3);
    \vertex[right=1cm of a3] (a4) {\(b\)};
    \vertex[right=1cm of a4] (a5);
    \vertex[right=2cm of a5] (a6) {\(u\)};

    \vertex[below=2em of a1] (b1) {\(d\)};
    \vertex[right=1cm of b1] (b2);
    \vertex[right=1cm of b2] (b3);
    \vertex[right=1cm of b3] (b4) {\(\overline d\)};
    \vertex[below=2em of a6] (b5) {\(\overline d\)};

    \vertex[above=of a6] (c1) {\(\overline u\)};
    \vertex[above=2em of c1] (c3) {\(d\)};
    \vertex at ($(c1)!0.5!(c3) - (1cm, 0)$) (c2);

    \diagram* {
      {[edges=fermion]
        (b1) -- (b2) -- (a2) -- (a1),
        (b5) -- (b4) -- (b3) -- (a3) -- (a4) -- (a5) -- (a6),
      },
      (a2) -- [boson, edge label=\(W\)] (a3),
      (b2) -- [boson, edge label'=\(W\)] (b3),

      (c1) -- [fermion, out=180, in=-45] (c2) -- [fermion, out=45, in=180] (c3),
      (a5) -- [boson, bend left, edge label=\(W^{-}\)] (c2),
    };

    \draw [decoration={brace}, decorate] (b1.south west) -- (a1.north west)
          node [pos=0.5, left] {\(B^{0}\)};
    \draw [decoration={brace}, decorate] (c3.north east) -- (c1.south east)
          node [pos=0.5, right] {\(\pi^{-}\)};
    \draw [decoration={brace}, decorate] (a6.north east) -- (b5.south east)
          node [pos=0.5, right] {\(\pi^{+}\)};
  \end{feynman}
\end{tikzpicture}
```

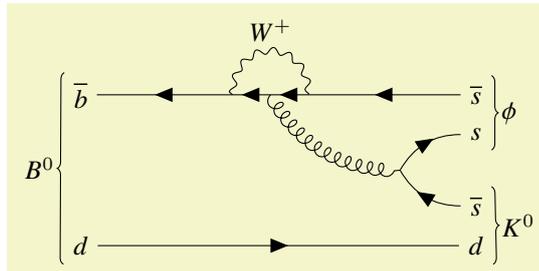

```
\begin{tikzpicture}
  \begin{feynman}
    \vertex (a1) {\(\overline b\)};
    \vertex[right=2cm of a1] (a2);
    \vertex[right=0.5cm of a2] (a3);
    \vertex[right=0.5cm of a3] (a4);
    \vertex[right=2cm of a4] (a5) {\(\overline s\)};

    \vertex[below=2cm of a1] (b1) {\(d\)};
    \vertex[below=2cm of a5] (b2) {\(d\)};

    \vertex[below=1.5em of a5] (c1) {\(s\)};
    \vertex[above=1.5em of b2] (c3) {\(\overline s\)};
    \vertex at ($(c1)!0.5!(c3) - (1cm, 0)$) (c2);

    \diagram* {
      {[edges=fermion]
        (a5) -- (a4) -- (a3) -- (a2) -- (a1),
      },
      (b1) -- [fermion] (b2),
      (c3) -- [fermion, out=180, in=-60] (c2) -- [fermion, out=60, in=180] (c1),
      (a3) -- [gluon, bend right] (c2),
      (a4) -- [boson, out=90, in=90, looseness=2.0, edge label'=\(W^{+}\)] (a2)
    };

    \draw [decoration={brace}, decorate] (b1.south west) -- (a1.north west)
          node [pos=0.5, left] {\(B^{0}\)};
    \draw [decoration={brace}, decorate] (a5.north east) -- (c1.south east)
          node [pos=0.5, right] {\(\phi\)};
    \draw [decoration={brace}, decorate] (c3.north east) -- (b2.south east)
          node [pos=0.5, right] {\(K^{0}\)};
  \end{feynman}
\end{tikzpicture}
```

# Index

This index only contains automatically generated entries. A good index should also contain carefully selected keywords. This index is not a good index.

## Acknowledgements

The original proof-of-concept for using TikZ to draw Feynman diagrams was done by the user 'Jake' on the TeX StackExchange. His original answer can be viewed at: `http://tex.stackexchange.com/a/87395/26980`.

I must also thank all the people who have used the development versions of TikZ-Feynman and offered suggestions to improve it and found bugs for me to fix.